\documentclass{aastex}          
\usepackage{spr-astr-addons}    
\usepackage{url}

\begin{document}

\title{Evidence of the Galactic outer ring $R_1R_2'$ from young open
clusters and OB-associations}

\shorttitle{Evidence of the Galactic outer ring from young open
clusters}

\shortauthors{Mel'nik et al.}

\author{A.~M. Mel'nik\altaffilmark{1}} \and
\author{P. Rautiainen\altaffilmark{2}} \and
\author{E.~V. Glushkova\altaffilmark{1}} \and
\author{A.~K. Dambis\altaffilmark{1}}

\altaffiltext{1}{Sternberg Astronomical Institute, Lomonosov
Moscow State University, Universitetskij pr. 13, Moscow 119991,
Russia}

\altaffiltext{2}{Department of Astronomy and Space Physics,
University of Oulu, P.O. Box 3000, FI-90014 Oulun yliopisto,
Finland}

\altaffiltext{}{e-mail: anna@sai.msu.ru}

\begin{abstract}

The distribution of young open clusters in the Galactic plane
within 3 kpc from the Sun suggests the existence of the outer ring
$R_1R_2'$ in the Galaxy. The optimum value of the solar position
angle with respect to the major axis of the bar,
$\theta_\textrm{b}$, providing the best agreement between the
distribution of open clusters and model particles   is
$\theta_\textrm{b}=35\pm10^\circ$. The kinematical features
obtained for young open clusters and OB-associations with negative
Galactocentric radial velocity $V_R$ indicate the solar location
near the descending segment of the outer ring $R_2$.

\end{abstract}


\maketitle

\newpage
\begin{figure*} \centering
\resizebox{12.0 cm}{!}{\includegraphics{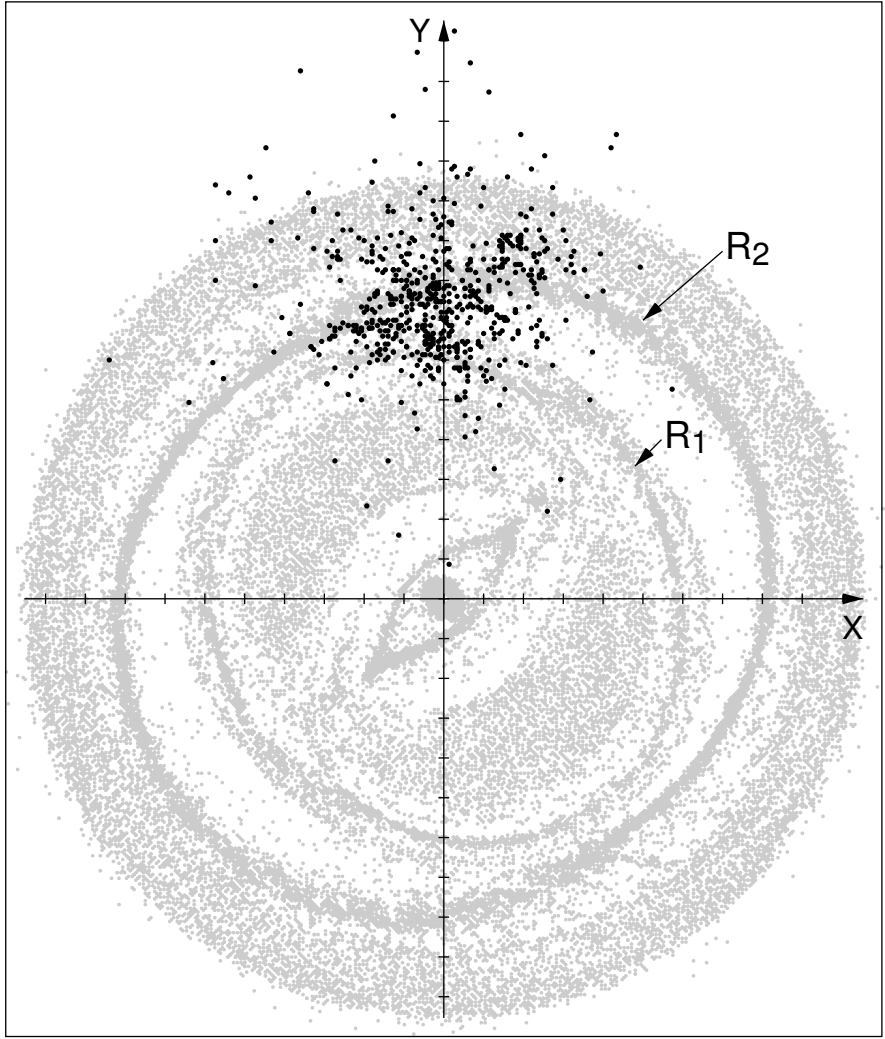}}
\caption{Distribution of young open clusters  (black circles) from
the catalog by \citet{dias2002} and model particles (gray circles)
in the Galactic plane. Only clusters with $\log \textrm{age}
<8.00$ and located within 0.5 kpc ($|z|< 0.5$ kpc) from the
Galactic plane are considered. The positions of model particles
(gas and OB particles) correspond to the position angle of the Sun
with respect to the bar of $\theta_\textrm{b}=45^\circ$. The
$X$-axis points in the direction of Galactic rotation and the
$Y$-axis is directed away from the Galactic center. One tick
interval along the $X$- and $Y$-axis corresponds to 1 kpc. The Sun
is located at (0, 7.5 kpc).  The arrows show the positions of the
outer rings $R_1$ and $R_2$. Of the two rings, $R_1$ is located a
bit closer to the Galactic center than the $R_2$.}
\label{distrib_total}
\end{figure*}

\newpage
\begin{figure*} \centering
\resizebox{\hsize}{!}{\includegraphics{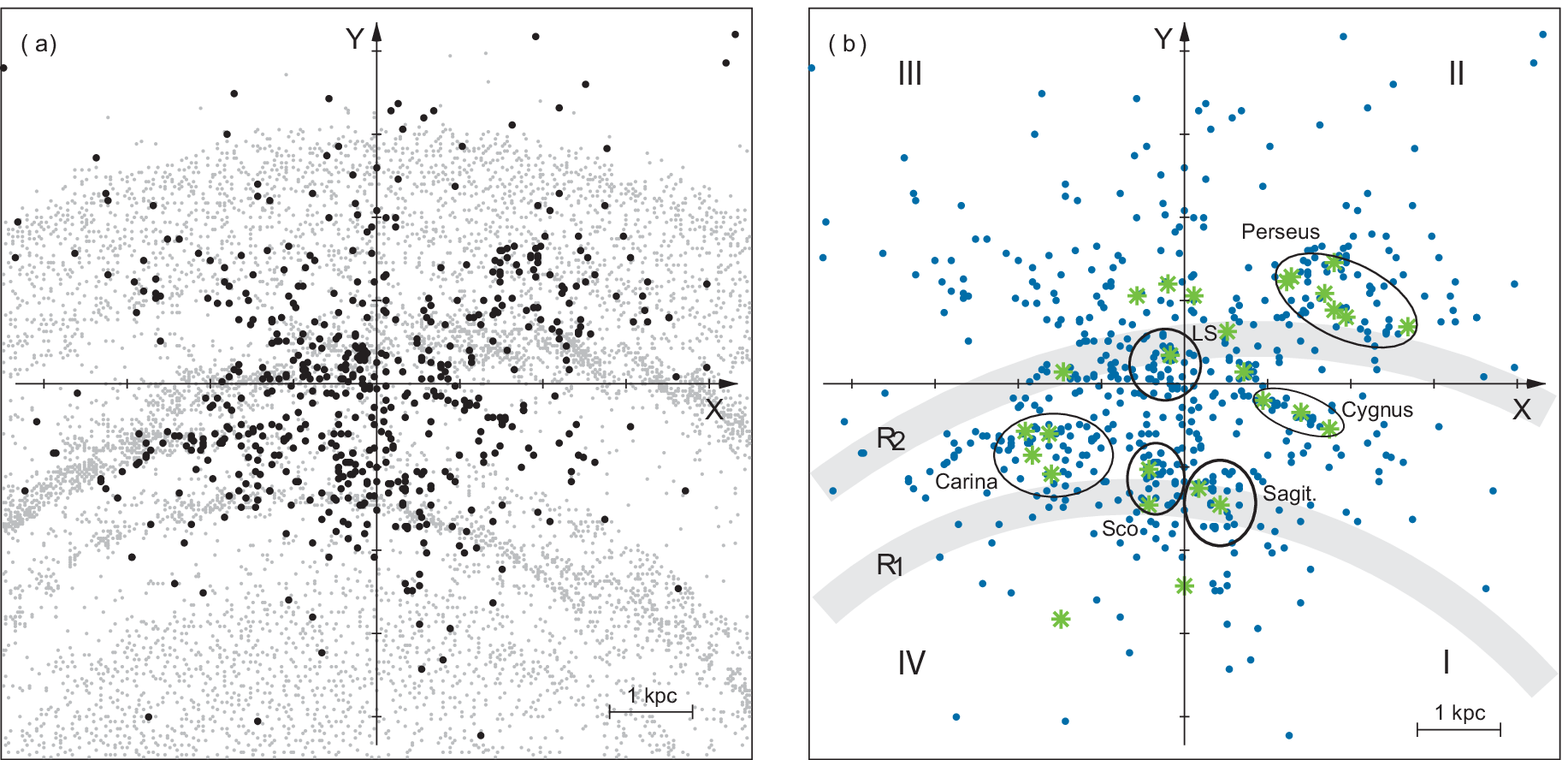}} \caption{(a)
Distribution of young ($\log \textrm{age} <8.00$) open clusters
(black circles) from the catalog by \citet{dias2002} and model
particles (grey points) in the Galactic plane zoomed in to a
larger scale.  The Sun is at the origin. The positions of the
model particles are drawn for $\theta_\textrm{b}=45^\circ$. The
$X$-axis points in the direction of Galactic rotation and the
$Y$-axis is directed away from the Galactic center. (b) The
distribution of young open clusters (circles colored blue in
electronic edition) and rich OB-associations (asterisks colored
green in electronic edition) in the Galactic plane. Only
OB-associations containing more than 30 members ($N_t>30$) in the
catalogue by \citet{blahahumphreys1989} are shown. The locations
of the outer rings $R_1$ and $R_2$ are indicated by gray arches.
The positions of the Sagittarius, Scorpio, Carina, Cygnus, Local
System (LS) and Perseus stellar-gas complexes  are drawn by
ellipses.   The Sagittarius and Scorpio complexes are located in
the vicinity of the ring $R_1$. The Perseus complex and  Local
System lie near  the ring $R_2$. The Carina complex is situated
in-between the two outer rings, where they seem to fuse together.
As for the Cygnus complex, its connection with some global
structure  is unclear.  Roman numerals show the numbers of
quadrants.} \label{distrib_local}
\end{figure*}

\section{Introduction}

Open clusters are compact groups of  stars born inside one giant
molecular cloud during a  short time interval. Young open clusters
are gravitationally bound objects  in distinction from
OB-associations,  which are loose  groups of O and B-type stars.
Such differences between young clusters  and OB-associations are
based on the comparison of their mass with the velocity
dispersions inside them.  The fact that all stars inside an open
cluster have nearly the same age  gives researchers the
opportunity to fit the cluster main sequence and colour-colour
diagrams  to  model  grids derived from zero age main sequence
(ZAMS) and a set of isochrones  corresponding to different
abundances. The result of fitting is  the determination of many
important physical characteristics of clusters, such as
heliocentric distance, age, and metallicity \citep{kholopov1980,
mermilliod1981}.

Young open clusters indicate the positions of giant molecular
clouds but  unlike gaseous objects,   open clusters  allow their
distances to be determined quite precisely with an accuracy of
$\sim 5\%$  as far as we ignore possible errors in the zero point
of the adopted ZAMS \citep{dambis1999}. So the concentration of
young open clusters in some complexes  suggests the presence of
gas there and traces the positions of spiral arms and Galactic
rings.

We suppose that the Galaxy contains a two-com-ponent outer ring
$R_1R_2'$ made up of  two elliptical gaseous rings stretched
perpendicularly to each other and located near the solar circle
(Fig.~\ref{distrib_total}). The sign of apostrophe means the
pseudoring $R_2'$ -- incomplete ring made up of two tightly wound
spiral arms. The idea that the Galaxy contains  outer rings was
first put forward  by \citet{kalnajs1991}.

Two main classes of outer rings and pseudorings  have been
identified: rings $R_1$ (pseudorings $R'_1$) elongated
perpendicular to the bar and rings $R_2$ (pseudorings $R'_2$)
elongated parallel to the bar. In addition, there is a combined
morphological type $R_1R_2'$ which exhibits elements of both
classes \citep{buta1995, buta1996, buta1991}.  Modelling shows
that outer rings are usually located near the Outer Lindblad
resonance (OLR) of the bar \citep[][and other papers]{schwarz1981,
byrd1994, rautiainen1999, rautiainen2000}.

\citet{comeron2014}  used the data from mid-infrared survey
\citep[Spitzer Survey of Stellar Structure in
Galaxies,][]{sheth2010}  to find that the frequency of  outer
rings is 16\% for all spiral galaxies located inside 20 Mpc and
over 40\% for disk galaxies of early morphological types (galaxies
with large bulges).  The above authors have also found that the
frequency of outer rings increases from $15\pm 2\%$ to $32\pm 7\%$
when going through the family sequence from SA to SAB, and
decreases again to $20\pm2\%$ for SB galaxies.

Note that the catalogue by Buta (1995) includes several tens of
galaxies with rings $R_1R_2'$.  Here are some examples of galaxies
with  the $R_1R_2'$ morphology that can be viewed  as possible
prototypes of the  Milky Way: ESO 245-1, NGC 1079, NGC 1211, NGC
3081, NGC 5101, NGC 5701,  NGC 6782, and NGC 7098. Their images
can be found in de Vaucouleurs Atlas of Galaxies by
\citet{buta2007} at http://bama.ua.edu/~rbuta/devatlas/

There is extensive  evidence  for the existence of the bar in the
Galaxy derived on the basis of infra-red observations
\citep{blitz1991, benjamin2005, cabrera-lavers2007,
gonzalez2012,churchwell2009} and gas kinematics in the central
region \citep{binney1991, englmaier1999, weiner1999}. The general
consensus is that  the major axis of the bar is oriented in the
direction $\theta_\textrm{b}=15\textrm{--}45^\circ$ in such a way
that the end of the bar closest to the Sun lies in  quadrant I,
where $\theta_\textrm{b}$ is  the position angle between the line
connecting the Sun and the Galactic center and the direction of
the major axis of the bar. The semi-major axis of the Galactic bar
is supposed to lie in the range $a=3.5\textrm{--}5.0$ kpc.
Assuming that its end is located close to its corotation radius
(CR), i.e. we are dealing with a so-called fast bar
\citep{debattista2000}, and that the rotation curve is flat, we
can estimate the bar angular speed $\Omega_b$, which appears to be
constrained to the interval $\Omega_b=40\textrm{--}65$ km s$^{-1}$
kpc$^{-1}$. This means that the OLR of the bar is located in the
solar vicinity: $|R_{OLR}-R_0|<1.5$ kpc. Studies of the kinematics
of old disk stars in the nearest solar neighbourhood, $r<250$ pc,
reveal the bimodal structure of the distribution of ($u$, $v$)
velocities, which is also interpreted to be a result of the solar
location near the OLR of the bar \citep[][and other
papers]{dehnen2000, fux2001}.

The explanation of the kinematics of young objects in the Perseus
stellar-gas complex (see its location  in
Fig.~\ref{distrib_local}b)  is a serious test for different
concepts of the Galactic spiral structure.  The  fact that  the
velocities of young stars in the Perseus stellar-gas complex are
directed toward the Galactic center, if interpreted in terms of
the density-wave concept \citep{lin1969}, indicates that the
trailing fragment of the Perseus arm must be located inside the
corotation circle (CR) \citep{burton1974,
melnik2001,melnik2003,sitnik2003}, and hence imposes an upper
limit for its pattern speed $\Omega_{sp}<25$ km s$^{-1}$
kpc$^{-1}$, which is inconsistent with the pattern speed of the
bar $\Omega_b=40\textrm{--}65$ km s$^{-1}$ kpc$^{-1}$ mentioned
above.

The studies of Galactic spiral structure  are usually  based on
the classical model developed by \citet{georgelin1976}, which
includes  four spiral arms with   a pitch angle of $\sim 12^\circ$
\citep[see e.g. \ the review by][]{vallee2013}. The main
achievement of this purely spiral model is that it can explain the
distribution of HII regions in the Galactic disk
\citep{russeil2003}.  This model became more physical after
incorporation of the bar into it \citep{englmaier1999}. However,
the bar and  spiral arms connected with it rotate with the angular
speed $\Omega_b=50\textrm{--}60$ km s$^{-1}$ kpc$^{-1}$, and this
model cannot explain the kinematics of young stars in the Perseus
complex. \citet{bissantz2003} developed  the model of
\citet{englmaier1999} by adding   a pair of spiral arms rotating
slower than the bar with $\Omega_{sp}=20$ km s$^{-1}$ kpc$^{-1}$.
However, it remains unclear what  mechanism can sustain this
slower spiral pattern in the disk.

\citet{liszt1985}   criticizes the use of kinematical distances
for tracing the Galactic spiral structure. He shows that
kinematical distances derived for HII regions, HI and CO clouds
can be wrong due to kinematic-distance ambiguity and velocity
perturbation from spiral arms. Moreover, \citet{adler1992} show
that bright spots in the diagrams (l, $V_{LSR}$) which are
interpreted as "clouds" can consist of a chain of clouds extending
over several kpc along the line of sight.

Models of the Galaxy with the outer ring $R_1R_2'$   reproduce
well the radial and azimuthal components of the residual
velocities (observed velocities minus the velocity due to the
rotation curve and  solar motion to the apex) of OB-associations
in the Sagittarius (see its location  in
Fig.~\ref{distrib_local}b) and Perseus complexes. The radial
velocities of most OB-associations in the Perseus stellar-gas
complex are directed toward the Galactic center and this indicates
the presence of the ring $R_2$ in the Galaxy, while the radial
velocities in the Sagittarius complex are directed away from the
Galactic center suggesting the existence of the ring $R_1$. The
nearly zero azimuthal component of the residual velocity of most
OB-associations in the Sagittarius complex precisely constrains
the solar position angle with respect to the bar major axis,
$\theta_\textrm{b}=45\pm5^\circ$. We considered models with
analytical bars and N-body simulations
\citep{melnikrautiainen2009, rautiainen2010}.

The classical model of Galactic spiral structure  can explain the
existence of so-called tangential directions related to the maxima
in the thermal radio continuum  as well as HI and CO emission,
which are associated with the tangents to the spiral arms
\citep{englmaier1999, vallee2008}. Models of a two-component outer
ring  can  also explain the existence of some of the tangential
directions  which, in this case, can be associated with the
tangents to the outer and inner rings. Our model diagrams (l
,V$_\textrm{LSR}$) reproduce the maxima in the direction of the
Carina, Crux (Centaurus), Norma, and Sagittarius arms.
Additionally, N-body model yields  maxima in the directions of the
Scutum and 3-kpc arms \citep{melnik2011, melnik2013}.

\citet{pettitt2014} simulated the (l,V$_\textrm{LSR}$) diagrams
for models  with  analytical bar. Their gas disks form the
two-component outer rings $R_1R_2$   200--500 Myr after  the start
of the simulation. The above authors  found observations to agree
best with the model with the solar position angle of
$\theta_\textrm{b}\approx 45^\circ$ and the bar pattern speed in
the range of $\Omega_b=50\textrm{--}60$ km s$^{-1}$ kpc$^{-1}$.

Elliptic outer rings can be divided into the ascending and
descending segments: in the ascending segments  galactocentric
distance $R$ decreases with  increasing azimuthal angle $\theta$,
which itself increases in the direction of galactic rotation,
whereas in the descending segments  distance $R$, on the contrary,
increases with increasing angle $\theta$. Ascending and descending
segments of the rings can be regarded as fragments of trailing and
leading spiral arms, respectively. Note that  if considered as
fragments of the spiral arms, the ascending segments of the outer
ring $R_2$ have the pitch angle of $\sim6^\circ$
\citep{melnik2011}.

\citet{schwarz1981}  associates   two main types of  outer rings
with two main families of periodic orbits existing near the OLR of
the bar \citep{contopoulos1980}. The main periodic orbits are
followed by numerous chaotic orbits, and this  guidance enables
elliptical rings to hold a lot of gas  in their vicinity. The
rings $R_1$ are supported by $x_1(2)$-orbits \citep[using the
nomenclature of][]{contopoulos1989} lying inside the OLR and
elongated perpendicular to the bar, while the rings $R_2$ are
supported by $x_1(1)$-orbits located slightly outside the OLR and
elongated along the bar.  However, the role of chaotic and
periodic orbits appears to be different inside and outside the CR
of the bar: chaos is dominant outside corotation, while most
orbits in the bar are ordered \citep{contopoulos2006, voglis2007}.
Not only periodic orbits induced by the bar and regular orbits
related to them, but also manifolds connected to the unstable
Lagrangian points near the ends of the bar, may contribute to the
formation of  outer rings and pseudorings \citep{romero-gomez2007,
harsoula2009, athanassoula2010}.

The study of classical Cepheids  from the catalogue by
\citet{berdnikov2000} revealed the existence of  "the
tuning-fork-like" structure in the distribution of Cepheids: at
longitudes $l>180^\circ$ (quadrants III and IV)  Cepheids
concentrate strongly to the arm located near the Carina complex
(the Carina arm), while at longitudes $l<180^\circ$ (quadrants I
and II) there are two regions of high surface density located near
the Perseus and Sagittarius complexes.  The term  "the Carina arm"
was used to designate the part of the Sagittarius-Carina arm
\citep[Fig. 11 in ][]{georgelin1976} that starts near the Carina
complex and continues to larger Galactocentric distances. In a
morphological study the Carina arm cannot be distinguished from
the ascending segment of the ring $R_2$. This morphology suggests
that outer rings $R_1$ and $R_2$ come closest to each other
somewhere near the Carina complex (see its location in
Fig.~\ref{distrib_local}b). We have also found some kinematical
features in the distribution of Cepheids, which suggest the
location of the Sun near the descending segment of the ring $R_2$
\citep{melnik2015}.

In  this  paper  we study the distribution and kinematics of young
open clusters and OB-associations. Section 2 describes the models
and catalogues used; Section 3  considers the morphological and
kinematical features that suggest the existence  of $R_1R_2'$ ring
in the Galaxy, and Section 4 presents the main conclusions.

\section{Catalogues and Models}

There are several large catalogues of open clusters.
\citet{dias2002}  compiled a  catalogue  of the physical and
kinematic parameters of open clusters using data reported by
different authors. This catalogue,   which is updated
continuously, is available at  http://www.astro.iag.usp.br/ocdb/
and presently lists 2167 clusters. \citet{kharchenko2013}
determined physical, structural and kinematic parameters of 3006
Galactic clusters. \citet{mermilliod1992} created   WEBDA database
of stars in open clusters (https://www.univie.ac.at/webda/), where
positional, photometric and spectroscopic data for individual
stars in cluster fields is stored. \citet{mermilliod2003} analysed
these data and derived the astrophysical parameters (reddening,
distance and age) of 573 open clusters.

In the last decade  many embedded clusters (stellar groups
recently born and still containing a lot of gas within their
volumes) were detected from near- and mid-infrared surveys
\citep[see e. g. \ the reviews by][]{glushkova2013, morales2013}.
However, distances to most of these objects reman uncertain mainly
because of  the variable extinction law in the field of embedded
clusters. The determination of their colour excesses requires
detailed photometric and spectroscopic studies. For example, the
estimates of the distance to the embedded cluster Westerlund 2
ranged from $r=2$ to 8 kpc, before a detailed study of this region
has been carried out \citep{carraro2013}. In this paper we
consider only optically observed clusters.

For our study we have chosen the catalogue by \citet{dias2002},
which provides the most reliable estimates of distances, ages and
other parameters. Its  new version (3.4) contains 627 young
clusters with the ages less than 100 Myr.

\citet{paunzen2006} established a list of 72 "standard" open
clusters covering a wide range of ages, reddenings and distances
selected on the basis of smallest errors from the available
parameters in the literature. Their analysis is based on the
averaged values from widely different methods and authors. The
authors then compared the derived mean values with the parameters
of open clusters published by \citet{dias2002}. They found that if
one uses the parameters of the catalogue by \citet{dias2002} then
the expected errors are comparable with those derived by averaging
the independent values from the literature. They concluded that
ages, reddenings and distances in the catalog by \citet{dias2002}
are good for statistical research.

We adopted the proper motions of open clusters based on the
Hipparcos catalogue \citep{hipparcos1997}  from the paper by
\citet{baumgardt2000}, and if they were absent there, from the
catalogues by \citet{glushkova1996, glushkova1997}, which are
available at https://www.univie.ac.at/webda/elena.html. In the
latter lists the proper motions were derived from the Four-Million
Star Catalogue of positions and proper motions
\citep[4M-catalogue,][]{volchkov1992} and then reduced to the
Hipparcos system.  We chose these catalogues of proper motions
because  of  the careful selection of star cluster members.

For kinematical study, we also  use OB-associations from the list
by \citet{blahahumphreys1989}, which includes 91 objects. Their
heliocentric distances $r_\textrm{BH}$ were reduced to the short
distance scale $r=0.8\cdot r_\textrm{BH}$ \citep{sitnik1996}. The
kinematical data were adopted from the catalogue by
\citet{melnikdambis2009}. The ages of OB-associations are supposed
to be less than 30 Myr \citep{humphreys1984,bressan2012}.

We use the simulation code developed by H. Salo
\citep{salo1991,salo2000}  to construct two different types of
models (models with analytical bars and models based on N-body
simulations), which reproduce the kinematics of OB-associations in
the Perseus and Sagittarius complexes. Among many models with
outer rings, we chose model 3 from the series of models with
analytical bars \citep{melnikrautiainen2009}  to  compare with
observations. This model has nearly flat rotation curve. The bar
semi-axes are equal to $a=4.0$~kpc and $b=1.3$~kpc. The positions
and velocities of $5\cdot 10^4$ model particles (gas+OB)   are
considered at time $T \approx 1$ Gyr from the start of the
simulation.  We  scaled and turned this model with respect to the
Sun to achieve the best agreement between  the velocities of model
particles and those of OB-associations in five stellar-gas
complexes identified by \citet{efremov1988}.

We adopt a solar Galactocentric distance of $R_0=7.5$ kpc
\citep{rastorguev1994, dambis1995, glushkova1998,
nikiforov2004,feast2008,groenewegen2008,reid2009b,dambis2013,
francis2014}. As model 3 was adjusted for $R_0=7.1$ kpc, we
rescaled all distances for model particles  by a factor of
$k=7.5/7.1$. Note that the particular choice of $R_0$ in the range
7–-9 kpc has practically no effect on the analysis of the
morphology and kinematics of stars located  within 3 kpc from the
Sun.

\section{Results}
\subsection{Space distribution of young open clusters}

The distribution of young open clusters in the Galactic plane can
reveal regions of intense star formation, which can be associated
with spiral arms or Galactic rings. Figure~\ref{distrib_total}
shows the distribution of young open clusters  from the catalog by
\citet{dias2002} and model particles in the Galactic plane. Only
clusters with ages less than 100 Myr and located within 0.5 kpc
($|z|<0.5$ kpc) from the Galactic plane are considered.  We can
see the model location of the outer rings $R_1$ and $R_2$
calculated for the solar position angle with respect to the bar
major axis of $\theta_\textrm{b}=45^\circ$.

Figure~\ref{distrib_local}a shows the distribution of young open
clusters  from the catalog by \citet{dias2002} and model particles
in a larger scale.  To avoid cluttering with other objects, we
made another plot (Fig.~\ref{distrib_local}b), where we indicate
the positions of rich OB-associations as well. The catalog by
\citet{blahahumphreys1989} includes 27 rich OB-associations
containing more than 30 members ($N_t>30$).
Figure~\ref{distrib_local}b also presents the positions of the
Sagittarius, Scorpio, Carina, Cygnus, Local System and Perseus
stellar-gas complexes from the list by \citet{efremov1988}.  We
can see a tuning-fork-like structure in the distribution of young
open clusters and OB-associations.  At negative x-coordinates (on
the left-hand side)  most of the clusters concentrate to the only
one arm   (the Carina arm), while at positive x-coordinates (on
the right-hand side) most of the clusters lie near the Perseus or
the Sagittarius  complexes. The Carina stellar-gas complex is
located in-between the two outer rings, where they come closest to
each other. The Sagittarius and Scorpio complexes lie near the
ring $R_1$, while the Perseus complex and Local System, on the
contrary, are situated near the ring $R_2$. In quadrant III, young
objects within $r<1.5$ kpc concentrate to the Sun, while more
distant objects distribute nearly randomly over a large area. As
for the Cygnus complex, its connection with some global structure
(outer ring or spiral arm) is unclear. However, the kinematics of
young objects in the Cygnus complex is similar to that in the
Perseus stellar-gas complex \citep{sitnik1996,sitnik1999},  and we
therefore tend to consider the Cygnus complex as a spur of the
ring $R_2$, which can be rather lumpy.

Using the model distribution, we can quite accurately approximate
the position of two outer rings by two ellipses oriented
perpendicular to each other. The outer ring $R_1$ can be
represented by the ellipse with the semi-axes $a_1=6.3$ and
$b_1=5.8$~kpc, while the outer ring $R_2$ fits well the ellipse
with $a_2=8.5$ and $b_2=7.6$~kpc. These values correspond to the
solar Galactocentric distance $R_0=7.5$ kpc. The ring $R_1$ is
stretched perpendicular to the bar and the ring $R_2$ is aligned
with the bar, hence the position of the sample of open clusters
with respect to the rings is determined by the position angle
$\theta_\textrm{b}$ of the Sun with respect to the major axis of
the bar.  The outer rings do not touch each other because gas
particles located on the orbits which cross near  the OLR were
scattered due to collisions during the  formation of the outer
rings. We now try to find the optimum angle $\theta_\textrm{b}$
providing the best agreement between the positions of the open
clusters  and the orientation of the outer rings.

\begin{figure*}
\resizebox{\hsize}{!}{\includegraphics{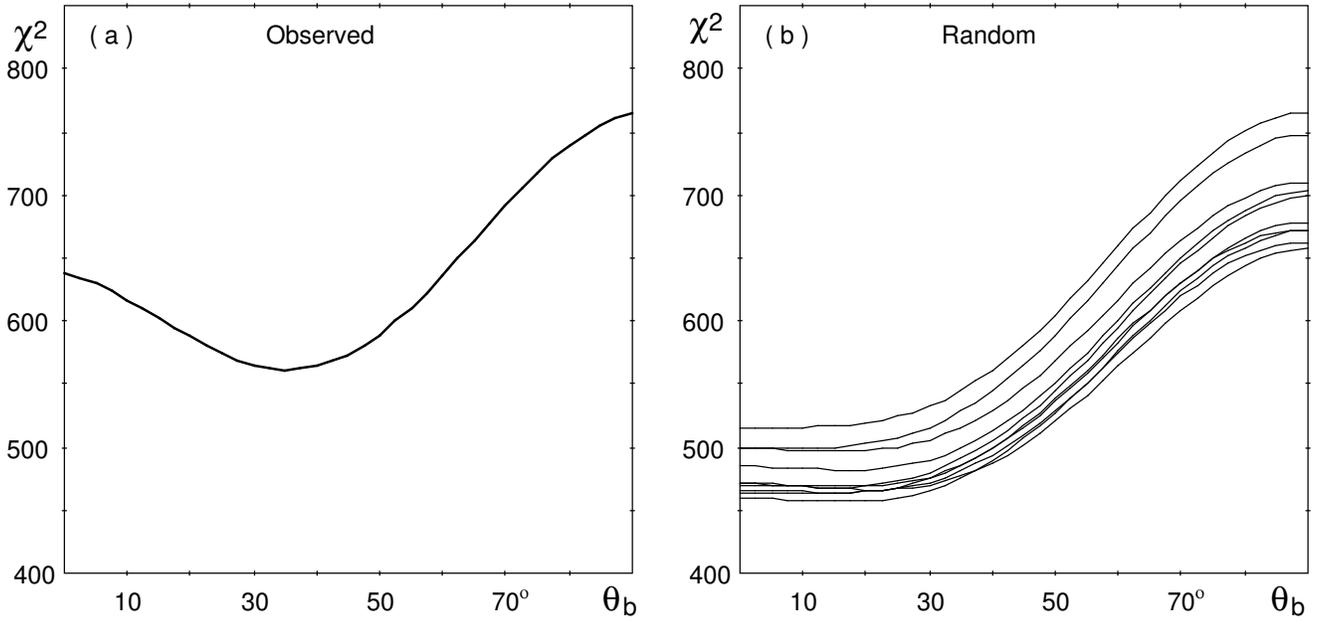}} \caption{The
$\chi^2$ functions calculated for different values of the solar
position angle $\theta_\textrm{b}$ with respect  to the major axis
of the bar. (a) The $\chi^2$ function derived for the distribution
of 564 young clusters from catalog by \citet{dias2002} located
within $r<3.5$ kpc of the Sun.  It  has a minimum at
$\theta_\textrm{b}=35\pm3^\circ$. (b) The $\chi^2$ functions
computed for 10 random samples containing 564 objects and
distributed in the Galactic plane in accordance with the power law
$n(r)\sim r^{-1}$ that simulates the effect of selection. We show
one such sample  in Fig.~\ref{ransam}a. The $\chi^2$ curves
calculated for random samples demonstrate a plateau at the
$0<\theta_\textrm{b} \le 30^\circ$ interval followed by a steep
rise in the $30<\theta_\textrm{b}<90^\circ$ interval. The
dissimilarity of the curves shown in "a" and "b" panels leads us
to conclude that the minimum of the curve in panel "a" is not due
to some model effects.} \label{chi2}
\end{figure*}

Figure~\ref{chi2} shows the  $\chi^2$ functions -- the sums of
normalized squared deviations  \citep{press1987} of open clusters
from the outer rings  -- calculated for different values of the
angle $\theta_\textrm{b}$. Figure~\ref{chi2}a shows the $\chi^2$
curve derived for the distribution of 564 young clusters   located
within $r< 3.5$ kpc of the Sun.  We can see a minimum  at
$\theta_{\textrm{min}}=35^\circ$ here. The random error of this
estimate is of about $\pm3^\circ$. Table~\ref{chi2obs} lists the
parameters of the observed sample: the number $N$ of clusters, the
standard deviation $\sigma$ of a cluster from the model position
of the outer rings, and the angle $\theta_{\textrm{min}}$
corresponding to the minimum on the $\chi^2$ curve.

\begin{figure*}
\resizebox{\hsize}{!}{\includegraphics{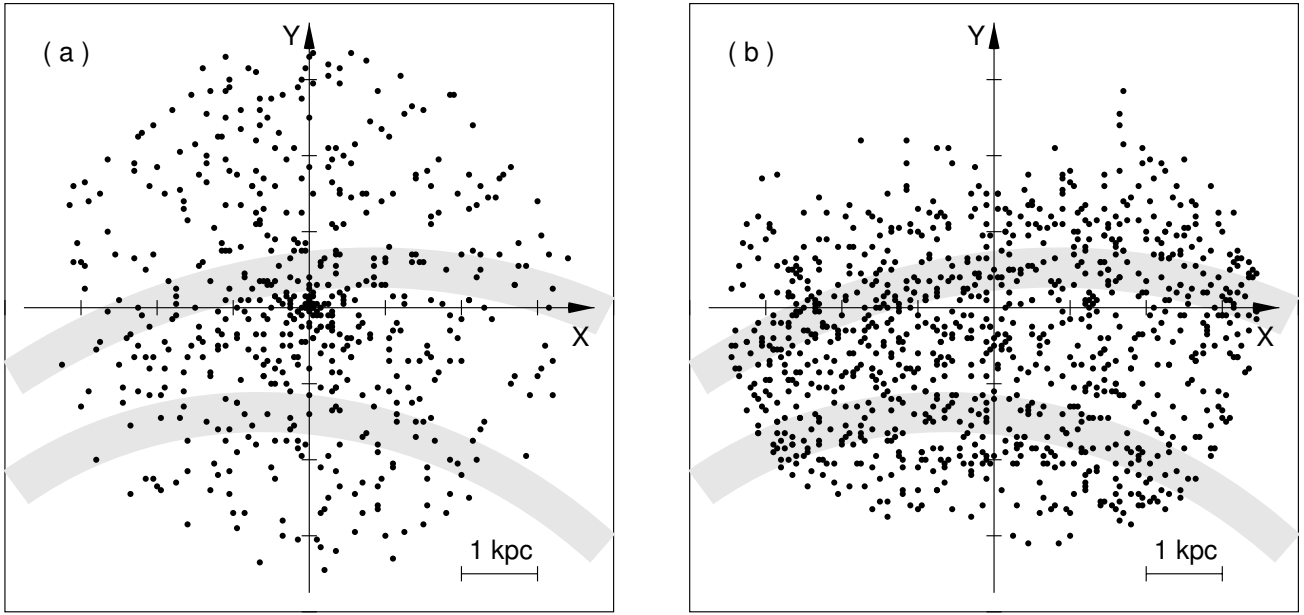}}
\caption{Examples of random samples generated to study selection
effects. (a) Simulated objects are distributed in the Galactic
plane in accordance with the power law $n(r)\sim r^{-1}$.  The
sample contains 564 objects. (b) Simulated objects are distributed
near the outer rings at the Gaussian law with the standard
deviation of $\sigma_r=0.8$ kpc. The ring $R_2$ is supposed to
contain 64\% of all objects. Only 20\% of $N_\textrm{mod}=5000$
objects  are shown. All simulated objects are located within 3.5
kpc from the Sun. The $X$-axis points in the direction of Galactic
rotation and the $Y$-axis is directed away from the Galactic
center.  The Sun is at the origin. } \label{ransam}
\end{figure*}

Figure~\ref{chi2}b shows  the $\chi^2$ functions computed for 10
random samples  also containing 564 objects and distributed along
the heliocentric distance $r$ in accordance with a  power law
$n(r)=r^{-1}$ that simulates the effect of selection (see section
3.2). An example of a such a random sample is shown in
Figure~\ref{ransam}a. The $\chi^2$ functions calculated for random
samples demonstrate a plateau in the $0<\theta_\textrm{b} \le
30^\circ$ range and a steep rise in the
$30<\theta_\textrm{b}<90^\circ$ interval (Fig.~\ref{chi2}b). A
comparison of images shown in Figure~\ref{chi2}a and
Figure~\ref{chi2}b indicates that the minimum at
$\theta_{\textrm{min}}=35^\circ$ exists only on the curve obtained
for the observed objects suggesting that it is not due to some
specific best-fitting angles which are, in principle, possible in
the cases involving only  a small region of the Galactic plane.

\begin{table}
\caption{Parameters of the sample}
 \begin{tabular}{lccc}
 \\[-7pt] \hline\\[-7pt]
 Sample  & $N_\textrm{obs}$   &   $\sigma$ & $\theta_{\textrm{min}}$  \\
  \\[-7pt] \hline\\[-7pt]
  $0<r<3.5$ kpc  & 564 & 0.80 kpc & $35\pm3^\circ$  \\
\hline
\end{tabular}
\label{chi2obs}
\end{table}

\begin{figure}
\centering \resizebox{\hsize}{!}{\includegraphics{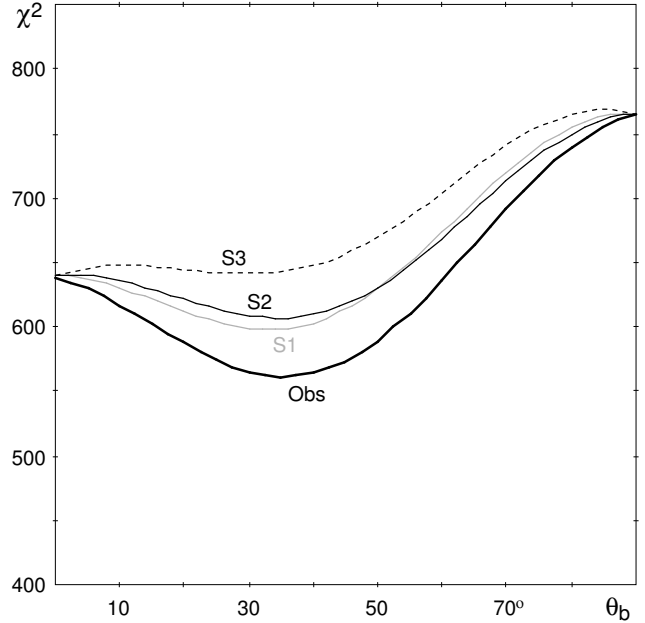}}
\caption{The $\chi^2$ functions computed for several samples of
objects designated as "Obs" (the thick solid line), "S1" (the gray
line),"S2" (the thin solid line), and "S3" (the dashed line). The
curve obtained for the observed sample of 564 young open clusters
from the catalog by \citet{dias2002} ("Obs") has the deepest
minimum. The modified sample "S1" is obtained from the observed
one by the mirror reflection of objects located in quadrants II
and III with respect to the axis $Y$, which  is identical with the
($l \to 360^\circ -l$) transformation. In  sample "S1" the
overdensity associated with the Perseus stellar-gas complex is
located in quadrant III. Sample "S2" is obtained by changing ($l
\to 360^\circ -l$) for objects located in quadrants IV and I,
while objects of quadrants I and II are left at their original
palaces. In  sample "S2" the overdensity associated with the
Carina complex is located in quadrant I between the outer rigs
$R_1$ and $R_2$. Sample "S3" is made by applying the ($l \to
360^\circ -l$) transformation to all objects. Here the
overdensities associated with the Perseus and Carina complexes are
located in quadrants III and I, respectively. We can see that the
$\chi^2$ curves calculated successively for the samples "Obs",
"S1", "S2", and "S3" have increasingly  shallow minima. }
\label{change}
\end{figure}

We made some modifications to the observed sample to show how the
overdensities in the Perseus and Carina complexes influence the
shape of the $\chi^2$ function. Figure~\ref{change} shows the
$\chi^2$ curves computed for the observed sample of open clusters
after the mirror reflection of some regions with respect to the
axis $Y$. All $\chi^2$ functions were calculated for the same
number of objects $N_\textrm{obs}=564$. Sample designated as "S1"
was obtained from the observed distribution by changing ($l \to
360^\circ -l$) for objects located in quadrants II and III. In
sample "S1" the overdensity associated with the Perseus complex is
located in quadrant III and the minimum of the corresponding
$\chi^2$ curve becomes shallower in comparison with that obtained
for the observed sample. This flattening of the $\chi^2$ curve is
caused by the fact that the ring $R_2$ reaches larger
$y$-coordinates in  quadrant II than in quadrant III
(Fig.~\ref{distrib_local}). It is true for all values of
$\theta_{\textrm{b}}$ from the expected interval 15--45$^\circ$.
Hence moving the Perseus complex into quadrant III increases
deviations from the rings, and, consequently, increases the
corresponding values of $\chi^2$.

Sample "S2" (Fig.~\ref{change}) is obtained by changing ($l \to
360^\circ -l$) for objects of quadrants IV and I, while objects of
quadrants I and II are left at their original places. In  sample
"S2" the overdensity associated with the Carina complex is located
in quadrant I just between the rings $R_1$ and $R_2$. This
transformation  increases the deviations from the rings for
objects of the Carina complex and causes the flattening the
$\chi^2$ curve as well.

Sample "S3" is a result of  the ($l \to 360^\circ -l$)
transformation of coordinates of all objects (Fig.~\ref{change}).
Here the overdensities associated with the Perseus and Carina
complexes lie in quadrants III and I, respectively. We can see
that the corresponding $\chi^2$ curve is practically flat in the
interval $\theta_{\textrm{b}}=15\textrm{--}45^\circ$. The
disappearance of the minimum here  is due to the tuning-fork-like
structure in the distribution of the observed objects. The mirror
reflection creates the tuning-fork-like structure pointed in the
opposite direction (one segment lies at positive $x$-coordinates
and two segments are located at negative $x$-coordinates), which
is inconsistent with the position of the outer rings obtained for
$\theta_\textrm{b}=15\textrm{--}45^\circ$.

\begin{figure}
\resizebox{\hsize}{!}{\includegraphics{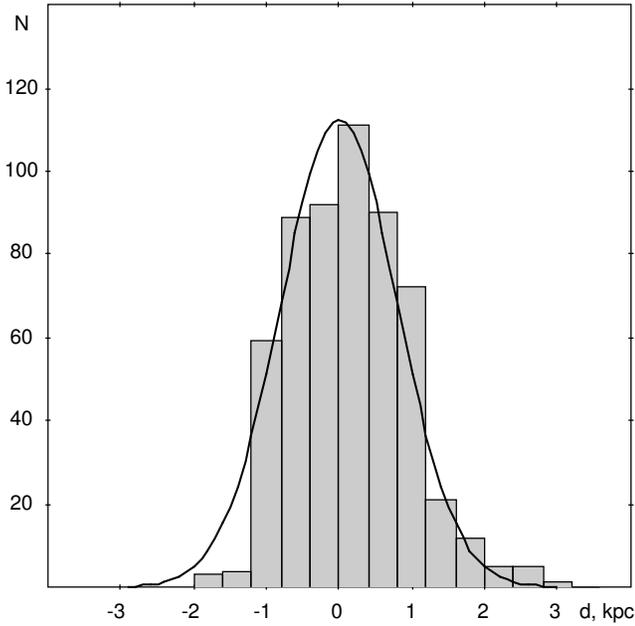}}
\caption{Distribution of the deviations (minimal distances) $d$ of
young open clusters  from the  model positions of the outer rings
calculated for $\theta_\textrm{b}=35^\circ$.  It can be
approximated by the Gaussian law with a standard deviation of
$\sigma_r=0.8$ kpc (the solid line).  Deviations from the rings in
the direction of increasing   $y$-coordinates are considered
positive and those in the opposite direction are considered
negative. The excess of positive deviations at $d>1.5$ kpc is due
to clusters located in the direction of the anticenter. The
fraction of clusters located within 1.5 kpc ($\sim2\sigma$) of the
rings appears to be 95\%, as expected in the case of the Gaussian
law.} \label{gauss}
\end{figure}

Figure~\ref{gauss} shows the histogram  of the deviations (minimal
distances) $d$ of young open clusters  from the model positions of
the outer rings calculated for $\theta_\textrm{b}=35^\circ$. Here
we can clearly see  the concentration of observed clusters to the
model positions of the outer rings. Deviations $d$ from the rings
in the direction of the increasing $y$-coordinates are considered
positive and  those in the opposite direction  are considered
negative. The distribution of deviations can be approximated by
the Gaussian law with a standard deviation of $\sigma=0.8$ kpc.
The excess of positive deviations at $d>1.5$ kpc is due to
clusters located in the direction of the anticenter. The fraction
of clusters located in the vicinity of 1.5 kpc ($\sim2\sigma$)
from the rings appears to be 95\%, which is in good agreement with
the Gaussian law.

We studied the effect of objects located in different regions on
the location of the minimum of the $\chi^2$ curve. Removing the
clusters located within the 0.5-kpc region ($r<0.5$ kpc) decreases
$\theta_{\textrm{min}}$ from 35 to $30^\circ$. This shift appears
due to the fact that the outer ring $R_2$ passes through the solar
vicinity at angle $\theta_{\textrm{b}}=90^\circ$. So nearby
clusters "favour" greater  $\theta_\textrm{b}$ values. However,
there is also an opposite effect: excluding  distant clusters
located in the direction of the anticenter ($y>2.5$ kpc) from the
observed sample increases  $\theta_{\textrm{min}}$ from 35 to
$40^\circ$. These clusters ($y>2.5$ kpc) are located far from both
outer rings and therefore "favour" the $\theta_\textrm{b}=0^\circ$
value in the case of which the ring $R_2$ crosses the $Y$-axis at
maximum Galactocentric distance. We hence tend regard the
uncertainty of $\pm5^\circ$ as a random error of our method.

We made an attempt  to simulate the influence of selection effects
by assigning to every model object the probability $P$ of its
detection. Generally, the detection of a cluster depends on a lot
of things: the richness and angular size of a cluster, the number
of resolved individual members  and their visual brightness, the
surface density of field stars, and the amount of extinction along
the line of sight \citep{morales2013}.

Let us suppose that the probability of detection of a cluster is
determined mainly by the brightness of its stars. Then the
probability of cluster detection is a function of its apparent
distance modulus $DM$ which depends on the heliocentric distance
to the cluster $r$ and the extinction $A_V$ toward it:

\begin{equation}
 DM= 5\lg r +10 + A_V,
\end{equation}

\noindent where $r$ is in kpc.

The probability $P$ of detection  of some objects  is usually
assumed to be equal to unity within some region of parameters and
to be exponentially decreasing function beyond it. We can thus
write the probability $P(DM)$  in the following way:

\begin{equation}
 \begin{array}{cc}
 P(DM)= &
 \left \{
 \begin{array}{lcr}
 1 & \textrm{if} &  DM<DM_0 \\
 &&\\
 e^{-(DM-DM_0)/s_0} & \textrm{else,} & \\
 \end{array}
 \right.\\
 \end{array}
 \label{p}
\end{equation}

\noindent where the scale factor $s_0$ and  zero point $DM_0$ are
determined by fitting between the  distributions of the distance
moduli $DM$ of observed and model clusters.

To simulate the sample of clusters we adopted the value of the
solar position angle $\theta_\textrm{tru}$ and scattered
$N_\textrm{mod}=5000$ model objects with respect to the outer
rings $R_1$ and $R_2$ in accordance with the Gaussian law with the
standard deviation of $\sigma_r=0.8$ kpc within 3.5 kpc of the
Sun, as it is shown in Figure~\ref{ransam}b.

Note that among 564  young open clusters from the catalog by
\citet{dias2002} located within $r< 3.5$ kpc, 408 objects (~70\%)
appears to lie within 0.8 kpc from one of the two outer rings, of
those 262 (64\%) are located in the vicinity of the ring $R_2$. We
therefore distributed simulated  objects among two rings placing
64\% of all objects  in the ring $R_2$.

To calculate the distance modulus for a model object  we must
assign  to it some value of the extinction $A_V$. That has been
done in accordance with the extinction of observed young ($\log
\textrm{age} <8.00$) clusters located in the nearby region and
derived from their colour excess $A_V=3.1E_{B-V}$
\citep{cardelli1989}. For each model objects situated at  point
(x, y) we selected observed young clusters from the catalog by
\citet{dias2002} located within the radius of $r_e=0.25$ kpc from
the point (x, y), and calculated their average value of extinction
$\overline{A_V}$. If there are less than $n_e<10$ observed
clusters in the region of radius $r_e$ we successively increased
the radius to $r_e=0.50$, 0.75, and 1.00 kpc. The  radius
$r_e=1.0$ kpc is the largest  considered and the corresponding
region always includes at least $n_e \ge 2$ observed clusters.
Note that more than  93\% of model objects have extinction
estimates  $\overline{A_V}$ averaged over $n_e> 10$ observed young
clusters.

The  distance moduli $DM$ for model objects were calculated as
follows:

\begin{equation}
 DM= 5\lg r +10 + \overline{A_V}+ \eta,
\end{equation}

\noindent  where $\eta$ is the error of the estimated distance
modulus. We supposed that $\eta$ is distributed according to the
Gaussian law with standard deviation $\sigma_e$,  which is
proportional to  the average extinction $\overline{A_V}$ with some
factor $k_e$:

\begin{equation}
\sigma_e = k_e \cdot \overline{A_V},
\end{equation}

\noindent  which is also determined, along with $s_0$ and $DM_0$,
by  fitting  the distributions of distance moduli of model and
observed objects.

\begin{figure}
\resizebox{\hsize}{!}{\includegraphics{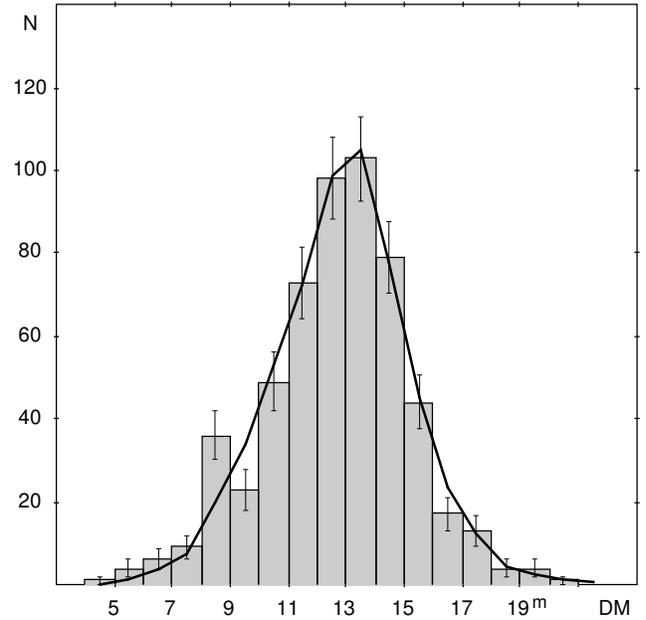}}
\caption{Distributions of the apparent  distance moduli $DM$ for
the observed and model clusters. The histogram shows the
distribution of 564 observed  clusters. The error bars show the
uncertainty due to Poisson noise $\sqrt{N_i^\textrm{obs}}$. The
solid line indicates one of the normalized distributions of model
objects obtained for the best-fitting parameters: $s_0=6.0^m$,
$k_e=0.55$, and $DM_0=9.0^m$.  } \label{his2}
\end{figure}

Figure~\ref{his2} shows the  distributions of the  distance moduli
$DM$ of observed and model clusters. The model distribution is one
of the best-fitting obtained with the parameters: $s_0=6.0^m$,
$k_e=0.55$, and $DM_0=9.0^m$. The difference in calculating the
model and observed distributions is that every model cluster  is
counted  with the weight factor  $P(DM)$, but not as unit entity
as it was done for the observed distribution. We then normalize
the model distribution so that it would contain the same number of
clusters as the observed sample. Thus  the normalized number of
model clusters $N_i^\textrm{mod}$ in the bin $DM_{i-1}$-- $DM_{i}$
is determined as follows:

\begin{equation}
 N_i^\textrm{mod}=\sum_{k=1}^{k=j} P_k(DM) \cdot N_\textrm{obs} / N_\textrm{mod},
\end{equation}

\noindent where $j$ is the number of model clusters  in the
distance-modulus bin considered, $P_k(DM)$ is the probability of
their detection,  $N_\textrm{obs}=564$ and $N_\textrm{mod}=5000$
are the numbers of observed and model clusters, respectively.

\begin{table}
\caption{Study of selection  effects }
 \begin{tabular}{lc}
 \\[-7pt] \hline\\[-7pt]
 $\theta_{\textrm{tru}}$   & $\theta_{\textrm{min}}$  \\
  \\[-7pt] \hline\\[-7pt]
 $32.0^\circ$  & $34.7\pm1.2^\circ$ \\
 $33.0^\circ$  & $35.5\pm1.2^\circ$ \\
 $34.0^\circ$  & $36.5\pm1.2^\circ$ \\
 $35.0^\circ$  & $37.5\pm1.2^\circ$ \\
 $36.0^\circ$  & $37.9\pm1.1^\circ$ \\
 $37.0^\circ$  & $38.8\pm1.1^\circ$ \\
 \hline
\end{tabular}
\label{chi2mod}
\end{table}

We computed the $\chi^2$ statistic for fitting the  distance
modulus distributions of the model and observed clusters
(Fig.~\ref{his2}) by the following formula:

\begin{equation}
 \chi^2=\sum_{i=1}^{i=17}\frac{(N_i^\textrm{mod}-
 N_i^\textrm{obs})^2}{N_i^\textrm{obs}},
\label{eq_his}
\end{equation}

\noindent where the scatter in each column of the histogram
(Fig.~\ref{his2}) is supposed to be due to the Poisson noise $\sim
\sqrt{N_i^\textrm{obs}}$ and the number of bins is $i=17$.

\begin{figure}
\resizebox{\hsize}{!}{\includegraphics{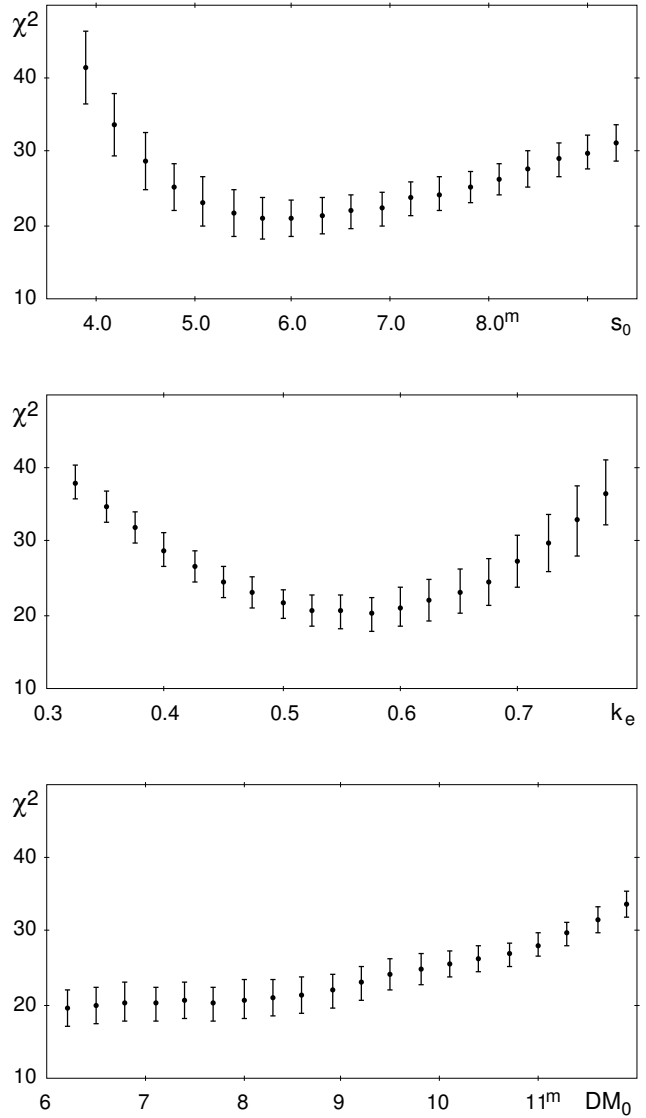}} \caption{The
behaviour of  $\chi^2$  (Eq.~\ref{eq_his}) as a function of one of
the parameters $s_0$, $k_e$, and $DM_0$ with all other parameters
fixed at their best-fitting values of $s_0=6.0^m$, $k_e=0.55$, and
$DM_0=9.0^m$. The error bars show the standard deviation in
determination of the $\chi^2$ values and allows us to estimate the
errors: $s_0=6.0\pm1.2^m$, $k_e=0.55\pm0.075$, and
$DM_0=9.0^{+0.8m}_{-3.0}$. The  parameter $DM_0$ is poorly
determined, but its $\chi^2$ function shows a kink near $\sim9^m$,
where initially flat behavior turns into appreciable growth.
Generally we can give only upper estimate for $DM_0$ equal to
$\sim 9^m$.} \label{min_par}
\end{figure}

Figure~\ref{min_par} shows  the $\chi^2$ functions calculated from
Eq.~\ref{eq_his}  plotted versus  one of the parameters $s_0$,
$k_e$, and $DM_0$ with all other fixed at their best-fitting
values of $s_0=6.0$, $k_e=0.55$, and $DM_0=9.0^m$. The  selection
zero point $DM_0$ is determined very poorly, but its $\chi^2$
function demonstrates a kink near  $\sim9^m$, where the
considerable growth follows the flat distribution. We can give
only  upper estimate of $DM_0$ equal to $\sim9^m$, which in the
case of zero extinction corresponds to $r_0=0.63$ kpc. Generally,
extinction must not be large near the Sun and its presence must
decrease the value of $r_0$ as well. Thus, the sample of young
open clusters from the catalog by \citet{dias2002} can be regarded
as complete only within the radius $r<0.63$ kpc.

Table~\ref{chi2mod} lists the  values $\theta_\textrm{tru}$ of the
angle $\theta_\textrm{b}$ used for simulating random samples and
the calculated values $\theta_{\textrm{min}}$ derived from the
location of the minimum of the $\chi^2$ curve. We determined the
$\theta_{\textrm{min}}$ values for 200 simulated samples and then
averaged them. A comparison of $\theta_{\textrm{tru}}$ and
$\theta_{\textrm{min}}$  reveals a small bias $\Delta_s$, so that
the calculated angles $\theta_{\textrm{min}}$ are always greater
than their true values. For $\theta_{\textrm{min}}=35^\circ$ this
bias  equals $\Delta_s=-2.5^\circ$.

However, if we suppose that  the probability of detection of model
clusters is  $P=1$ we obtain  a  systematical correction with the
opposite sign,  $\Delta_s=+5^\circ$, so that the calculated values
are always smaller than the true ones. Generally, this shift is
due to the objects located in the direction of the anticenter.

Note that the shift between the $\theta_{\textrm{tru}}$ and
$\theta_{\textrm{min}}$ values decreases with decreasing the
scatter $\sigma_r$  of model objects with respect to the outer
rings.

We tend  to regard the $\theta_\textrm{b}=35^\circ$ value as a
good compromise to account for the combined effect of different
factors. We summed the random $\pm5^\circ$ and systematical
$\pm5^\circ$ errors because of their possible correlation to
obtain   an upper limit of $\sim 10^\circ$ for the combined error.

Note that  our  study of the sample of classical Cepheids yields
$\theta_\textrm{b}=37\pm13^\circ$ for the position angle of the
Sun with respect to the bar's major axis \citep{melnik2015}. And
the cause of this coincidence is that a tuning-fork-like structure
is also present in the Cepheid distribution.

\subsection{Surface density and extinction in different sectors}

To study the distribution of clusters with respect to the Sun, we
calculated their surface density in the Galactic plane. The
variations in the surface density  and extinction toward clusters
provide some information about  selection effects and the
distribution of dust in the Galaxy. Here we  verify the hypothesis
that selection effects, if we consider their influence in
$45^\circ$-width sectors of the Galactic plane, affect the sample
of young clusters in nearly the same way irrespective of the
direction, implying that the density maxima in the Carina and
Perseus complexes can be attributed to real density enhancement
rather than  the lack of corresponding observations in other
sectors.

To study selection effects, we  subdivide the sample of young open
clusters into 8 subsamples confined by $45^\circ$-width sectors:
$l=0\textrm{--}45^\circ$, 45--90$^\circ$, 90--135$^\circ$,
135--180$^\circ$, 180--225$^\circ$, 225--270$^\circ$,
270--315$^\circ$, and 315--360$^\circ$. For each sector we
calculate the number of clusters in annuli of width $\Delta r=0.5$
kpc  in the Galactic plane. The ratio of the number of clusters in
1/8-th of an annulus to its area gives us the average surface
density of clusters in each sector per kpc$^2$. Table~\ref{sel}
lists the surface density values $n_1$, $n_2$,..., $n_8$ in
different sectors derived for different distance  intervals $r_1$
-- $r_2$. It also gives the total surface density $n_0$ averaged
over all sectors and the standard deviation $\sigma_0$ of the
surface densities $n_1$, $n_2$,..., $n_8$ in sectors 1--8 from the
average surface density $n_0$, which  depends only on distance.

\begin{figure*} \centering
\resizebox{16 cm}{!}{\includegraphics{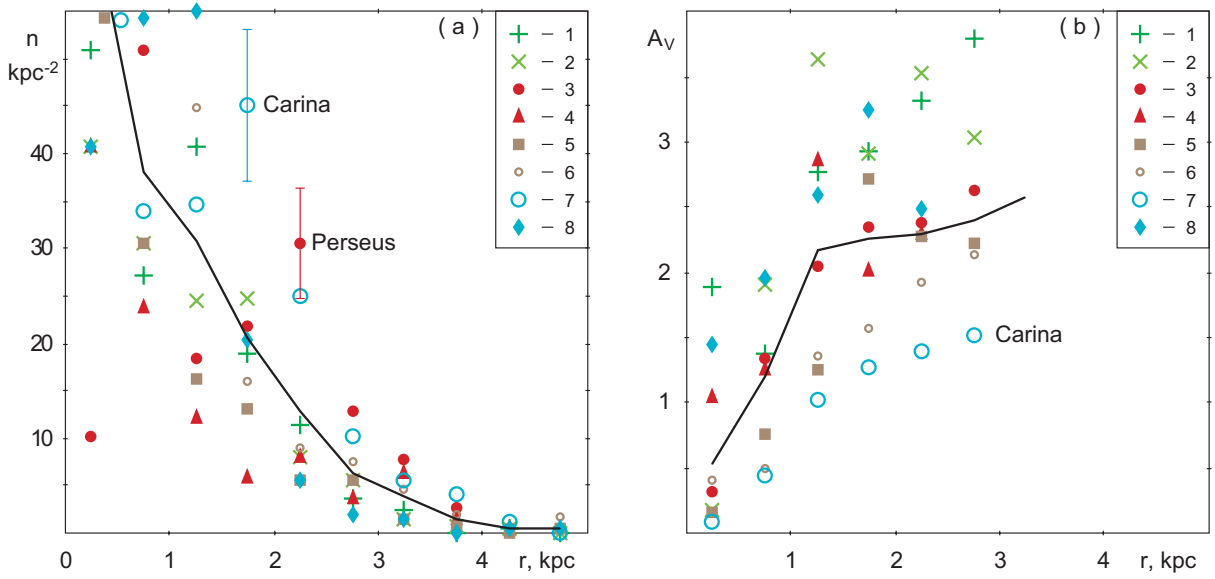}} \caption{(a)
Surface density $n$  in the Galactic plane calculated for young
open clusters as a function of  the  heliocentric distance $r$ for
8 different sectors.  Different symbols correspond to different
sectors. In electronic edition four quadrants are represented by
different colors.  The errors of the computed surface density
$n(r)$  in different sectors caused by the Poisson noise are, on
average, twice smaller than the corresponding  $\sigma_0$ values.
The individual error bars are shown only for the Carina and
Perseus complexes. The surface density $n_0$ averaged over all
sectors is shown by a solid line. (b) The extinction $A_V$ along
$r$ calculated for different sectors  and $A^0_{V}$ averaged over
all sectors (the solid line). The mean uncertainty of
sector-averaged $A_{V}$ values is 0.4$^m$. The numbers 1--8
indicate the sectors: $l=0\textrm{--}45^\circ$ (1), 45--90$^\circ$
(2), 90--135$^\circ$ (3), 135--180$^\circ$ (4), 180--225$^\circ$
(5), 225--270$^\circ$ (6), 270--315$^\circ$ (7), and
315--360$^\circ$ (8).} \label{surf_den_all}
\end{figure*}

Figure~\ref{surf_den_all}a  shows the variations of surface
densities $n_1$, $n_2$,..., $n_8$, and $n_0$ along heliocentric
distance $r$.  The density maxima corresponding to the Carina and
Perseus complexes are also indicated. The corresponding peaks of
surface density are underlined in Table~\ref{sel}. The  standard
deviation $\sigma_0$ in the distance interval 0.5--3 kpc amounts,
on average, to 40\% of the total surface density $n_0(r)$.  The
errors $\sigma_p$ of the estimated surface density $n(r)$ in
different sectors caused by the Poisson noise are, on average,
twice smaller than the corresponding standard deviation $\sigma_0$
arising from the scatter of values obtained for different sectors.
The average in the distance interval  1.5--2.5 kpc is
$\overline{\sigma_p}=4$ cluster$\cdot$kpc$^{-2}$. The peaks in the
Carina and Perseus complexes can be seen to deviate significantly
from the total surface density $n_0(r)$ -- the deviations amount
to (1.9--2.1)$\sigma_0$ or $3\sigma_p$. We can hence conclude that
these regions of enhanced density really exist at a significance
level of $P\ge 2\sigma_0$. We cannot say the same of the
Sagittarius complex (sector 1, $r=1\textrm{--}1.5$ kpc), where
surface density exceeds the average value only by $0.7 \sigma_0$
or $1.1\sigma_p$. However, this density peak extends beyond the
corresponding sector and continues into the adjacent sector
(sector 8, $r=1\textrm{--}1.5$ kpc) where  density exceeds the
average value by $1 \sigma_0$ or $1.2\sigma_p$ (Table~\ref{sel}).

\newpage

We  assume that surface density of clusters decreases due to
selection effects  at a power law  rate: $n(r)\sim r^{-\beta}$. We
use seven  surface density values in the interval 0--3.5 kpc to
determine exponent $\beta$ and its error $\varepsilon_{\beta}$ for
functions $n_1(r)$, $n_2(r)$,..., $n_8(r)$, and $n_0(r)$
(Table~\ref{sel}). The  exponent $\beta$ derived for  the total
distribution $n_0(r)$ appears to be $\beta=1.03\pm0.21$ and the
corresponding exponents for different sectors, as a rule, coincide
with this value within the errors. The  only exception is the
sector 180--225$^\circ$ (5), where $\beta=1.62\pm0.18$. The steep
decline of surface density in this sector is possibly due to the
physical lack of clusters in  quadrant III.

Figure~\ref{surf_den_all}b demonstrates the variations of
extinction $A_V$  toward open clusters along the distance $r$ in
different sectors  of the Galactic plane.  We calculated
extinction $A_V$   for each sector and each distance range  based
on the average colour excess of open clusters located there. Also
shown is the average extinction $A^0_{V}(r)$ calculated for all
sectors. The mean uncertainty of sector-averaged $A_{V}$ values is
0.4$^m$. Several interesting features are immediately apparent.
First, extinction grows most rapidly in the sectors 0--45$^\circ$
(1), 45--90$^\circ$ (2) and 315--360$^\circ$ (8). Second, the
minimal values of extinction are observed in the direction of the
Carina complex 270--315$^\circ$ (7) and in the direction of the
anticenter 225--270$^\circ$ (6). Third, variations of extinction
in the sector of the Perseus stellar-gas complex 90--135$^\circ$
(3) are very close to those of average extinction $A^0_{V}$.

Note that we  found similar features in variations of extinction
for classical Cepheids \citep{melnik2015}. Both groups of young
objects, open clusters and Cepheids, show  a sharp growth of
extinction in the direction toward the Galactic center \citep[see
also][]{neckel1980,marshall2006}. This steep increase can be
attributed to the presence of the Galactic outer ring $R_1$ at the
heliocentric distance 1--2 kpc  in this direction. The other
feature --  lower extinction in the direction of the Carina
complex 270--315$^\circ$ -- can also be seen in the sample of
classical Cepheids. Generally,  low extinction in some direction
means  that the line of sight crosses interstellar medium with
lower concentration of dust. Probably, there is a lack of dust in
the space between the Sun and the Carina complex. The same is true
for the anticenter direction.  Moreover, lack of dust often
correlates with lack of gas \citep[][and references
therein]{xu1997,amores2005}.

\begin{table*}
 \caption {Surface density in different sectors}
 \begin{tabular}{ccccccccccc}
  \hline
  $r_1$ -- $r_2$  &$n_1$&$n_2$&$n_3$&$n_4$&$n_5$&$n_6$&$n_7$&$n_8$&$n_0$&$\sigma_0$  \\
  &0--45$^\circ$&45--90$^\circ$&90--135$^\circ$&135--180$^\circ$&180--225$^\circ$&225--270$^\circ$&270--315$^\circ$&
 315--360$^\circ$&All& \\
  kpc&&&Perseus&&&&Carina&&&\\
   \\[-7pt] \hline\\[-7pt]
 0.0  --  0.5&  50.9& 40.7& 10.2& 40.7&173.2&122.2& 71.3& 40.7&   68.8& 53.3\\
 0.5  --  1.0&  27.2& 30.6& 50.9& 23.8& 30.6& 54.3& 34.0& 54.3&   38.2& 12.8\\
 1.0  --  1.5&  40.7& 24.4& 18.3& 12.2& 16.3& 44.8& 34.6& 55.0&   30.8& 15.3\\
 1.5  --  2.0&  18.9& 24.7& 21.8&  5.8& 13.1& 16.0& \underline{45.1}& 20.4&   20.7& 11.5\\
 2.0  --  2.5&  11.3&  7.9& \underline{30.6}&  7.9&  5.7&  9.1& 24.9&  5.7&   12.9&  9.5\\
 2.5  --  3.0&   3.7&  5.6& 13.0&  3.7&  5.6&  7.4& 10.2&  1.9&    6.4&  3.7\\
 3.0  --  3.5&   2.4&  1.6&  7.8&  6.3&  1.6&  4.7&  5.5&  1.6&    3.9&  2.5\\
   \\[-7pt] \hline\\[-7pt]
 $\beta$      & 1.08&   1.05&  0.97&  0.89&  1.62&  1.28&  0.79&  1.34&    1.03&  \\
 $\varepsilon_{\beta}$& $\pm0.34$&  $\pm0.35$&  $\pm0.36$&  $\pm0.15$&  $\pm0.18$&  $\pm0.19$&  $\pm0.27$&$\pm0.50$&$\pm0.21$&\\
    \\[-7pt] \hline\\[-7pt]
\multicolumn{11}{l}{ All values of surface density $n_1$,
$n_2$,..., $n_0$, $\sigma_0$ are  in the units of
cluster$\cdot$kpc$^{-2}$}
\end{tabular}
\label{sel}
\end{table*}

\subsection{Rotation curve}

 We use  the sample of young ($\log \textrm{age} <8.00$) open
clusters  from the catalog by \citet{dias2002} to determine the
parameters of the rotation curve and compare them with those
derived from the samples of OB-associations and classical
Cepheids. We selected open clusters located within 3 kpc from the
Sun and within 0.5 kpc ($|z|<0.5$ kpc) from the Galactic plane. In
total,  we have 187 young clusters with known proper motions,  and
212 with known line-of-sight velocities. The line-of-sight
velocities of clusters were taken from the catalogue by
\citet[][version 3.4]{dias2002}. Of them $\sim50\%$  were
determined by \citet{dias2014}. We  use only line-of-sight
velocities and proper motions determined from the data  for at
least two cluster members ($n_{vr}\ge2$, $n_{pm}\ge2$). Of 187
cluster proper motions 53 were adopted from the list by
\citet{baumgardt2000}  and the remaining 134 proper motions, from
the list by \citet{glushkova1996,glushkova1997},  which is
available at http://www.sai.msu.su/groups/cluster/cl/pm/. The
proper motions used in this study are derived or reduced to the
Hipparcos system \citep{hipparcos1997}.

We suppose that the motion of young objects in the disk obeys a
circular rotation law. Then  we can write the so-called Bottlinger
equations for the line-of-sight velocities $V_r$ and proper
motions along  Galactic longitude $\mu_l$:

\begin{equation}
\begin{array}{l}
V_r=R_0\cdot(\Omega-\Omega_0)\cdot\sin l \cdot\cos b  \\
\qquad -(u_0 \cdot \cos l \cdot \cos b+v_0\cdot\sin l \cdot\cos
b+w_0\cdot \sin b),
\end{array}
\end{equation}

\begin{equation}
\begin{array}{l}
4.74\cdot\mu_l\cdot r=R_0\cdot(\Omega-\Omega_0)\cdot\cos l - \Omega\cdot r \cdot\cos b \\
\qquad -(-u_0\cdot\sin l +v_0\cdot\cos l). \label{mu}
\end{array}
\end{equation}

The parameters $\Omega$ and $\Omega_0$ are the angular rotation
velocities  at the Galactocentric distance $R$ and at the distance
of the Sun $R_0$, respectively. The  velocity components $u_0$ and
$v_0$ characterize the solar motion with respect to the centroid
of objects considered in the direction toward the Galactic center
and Galactic rotation, respectively. The velocity component $w_0$
is directed along the $z$-coordinate and we set it equal to
$w_0=7.0$ km s$^{-1}$.  The factor 4.74 converts the left-hand
part of Eq.~\ref{mu} (where proper motion and distance are in mas
yr$^{-1}$ and kpc, respectively) into the units of km s$^{-1}$.

We expanded the angular rotation velocity $\Omega$  at
Galactocentric distance $R$ into a power series in $(R-R_0)$:

\begin{equation}
 \Omega= \Omega_0 + \Omega'_0\cdot(R-R_0) + 0.5\cdot\Omega''_0\cdot(R-R_0)^2,
\end{equation}

\noindent where $\Omega'_0$ and $\Omega''_0$ are its  first and
second derivatives taken at the solar Galactocentric distance.

We simultaneously solve the equations for the line-of-sight
velocities and proper motions.  We also use weight factors
$p_{vr}$ and $p_{vl}$ to balance the errors of line-of-sight and
transversal velocity components:

\begin{equation}
p_{vr} =(\sigma_0^2+\varepsilon^2_{vr})^{-1/2},
\end{equation}

\begin{equation}
p_{vl} =(\sigma_0^2+(4.74\cdot\varepsilon_{\mu l}\cdot
r)^2)^{-1/2},
\end{equation}

\noindent where  $\sigma_0$ is the so-called "cosmic" velocity
dispersion, which is approximately equal to the rms deviation of
the velocities from the rotation curve  \citep[for more details
see][]{dambis1995, melnik1999,melnikdambis2009}.  We adopted the
errors of the line-of-sight velocities $\varepsilon_{vr}$ and
proper motions $\varepsilon_{\mu l}$  from the corresponding
catalogues.

We used an iterative cycle with 3$\sigma$ clipping to determine
both the parameters of motion and the "cosmic" velocity dispersion
$\sigma_0$.  At each iteration the line-of-sight velocities that
deviate  by more than $3 \sigma_0$ from the computed rotation
curve  were eliminated. We also excluded proper motions which
deviate by more than 6.0 mas yr$^{-1}$ ($3\cdot2$ mas yr$^{-1}$,
where 2 mas yr$^{-1}$ is the average error of proper motions
considered) from the rotation curve. The rotation curve and solar
motion parameters remain practically unchanged in subsequent
iterations, but $\sigma_0$ decreases from 23 to 15 km s$^{-1}$.
The final sample includes 156 and 209 equations for proper motions
and line-of-sight velocities, respectively.

\begin{figure}
\resizebox{\hsize}{!}{\includegraphics{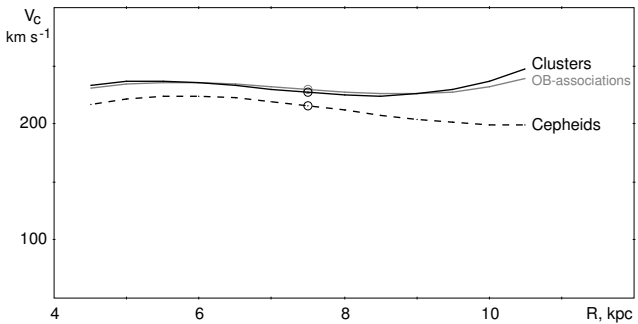}}
\caption{Galactic rotation curve derived from an analysis of
line-of-sight velocities and proper motions of young open clusters
(the black line), which  practically coincides with that obtained
for OB-associations (the gray line).  The rotation curve computed
for Cepheids lies systematically lower (the dashed line). The
circle shows position of the Sun.} \label{rot_curve}
\end{figure}

Table~\ref{rot} lists the final values of the parameters of the
rotation curve $\Omega_0$, $\Omega'_0$, $\Omega''_0$ and solar
motion $u_0$, $v_0$ and  the final value of velocity dispersion
$\sigma_0$ calculated for young open clusters. Also listed are the
number of conditional equations $N$ and the value of Oort constant
$A=-0.5R_0\Omega'$. For comparison we also give the parameters
derived for the sample of OB-associations \citep{melnikdambis2009}
and those inferred for classical Cepheids \citep{melnik2015}. We
can see a good agreement between the parameters obtained for young
open clusters and OB-associations (see also Fig.~\ref{rot_curve}).

Note the large value of $\Omega_0=30.3\pm1.2$ km s$^{-1}$
kpc$^{-1}$ obtained for young open clusters, which coincides with
that calculated for OB-associations and maser sources,
$\Omega_0=31\pm1$ km s$^{-1}$ kpc$^{-1}$
\citep{reid2009a,melnikdambis2009, bobylev2010}. On the other
hand, the angular velocity $\Omega_0$ at the solar Galactocentric
distance estimated from the kinematics of Cepheids
\citep{feast1997, bobylev2012} is systematically lower than the
value derived from young open clusters and OB-associations. The
parameters computed for young open clusters and Cepheids are
consistent within the errors except for the second derivative
$\Omega''_0$,  which is equal to $1.65\pm0.29$ and $1.07\pm0.17$
km s$^{-2}$ kpc$^{-1}$, respectively (Table~\ref{rot}). Moreover,
the rotation curve of Cepheids seems to be slightly descending,
whereas that of OB-associations is nearly flat within the 3 kpc.
The cause of this discrepancy  is currently not clear.

The standard deviation  of the velocities from the rotation curve
equals $\sigma_0=15.2$, 7.16 and 10.84 km s$^{-1}$ for young open
clusters, OB-associations and Cepheids, respectively
(Table~\ref{rot}), which is consistent with other estimates
\citep[e.g. \ ][]{zabolotskikh2002}. The larger value of
$\sigma_0=15.2$ km s$^{-1}$ derived for clusters reflects   the
quality of the kinematical data rather than the physical scatter:
all these objects are quite young, their ages do not exceed 200
Myr and they all must demonstrate the same  physical scatter with
respect to rotation curve to within 1--2 km s$^{-1}$. Possibly,
the crowding in  cluster fields prevents accurate measurements of
line-of-light velocities and proper motions. Furthermore, the
line-of-sight velocities of young open clusters are based on
measured velocities of early-type stars, which usually have low
accuracy because of the small number and large width of spectral
lines involved.

 \begin{table*}
 \centering
 \caption{Parameters of the rotation curve and solar motion}
 \begin{tabular}{lcccccccl}
  \\[-7pt] \hline\\[-7pt]
  Objects &$\Omega_0$ & $\Omega'_0$ & $\Omega''_0$ & $u_0$ & $v_0$ &  A & $\sigma_0$ & N \\
  & km s$^{-1}$  & km s$^{-1}$  & km s$^{-1}$  &
  km s$^{-1}$ & km s$^{-1}$  &km s$^{-1}$ &km s$^{-1}$ &   \\
  &   kpc$^{-1}$ & kpc$^{-2}$ & kpc$^{-3}$ &
   & & kpc$^{-1}$& &  \\
  \\[-7pt] \hline\\[-7pt]
 Young open clusters & 30.3 & -4.73 & 1.65 & 9.8 & 11.1 &  17.7 & 15.20 & 365\\
 $A<10^8$ yr     & $\pm1.2$ & $\pm0.24$ & $\pm0.29$ & $\pm1.3$ & $\pm1.5$  & $\pm0.9$ &\\
  \\[-7pt] \hline\\[-7pt]
 OB-associations & 30.6 & -4.73 & 1.43 & 7.7 & 11.6 &  17.7 & 7.16 & 132\\
 & $\pm0.9$ & $\pm0.18$ & $\pm0.21$ & $\pm1.0$ & $\pm1.3$  & $\pm0.7$ &\\
  \\[-7pt] \hline\\[-7pt]
 Classical Cepheids & 28.8 & -4.88 & 1.07 & 8.1 & 12.7 &  18.3 & 10.84 & 474\\
  & $\pm0.8$ & $\pm0.14$ & $\pm0.17$ & $\pm0.8$ & $\pm1.0$  & $\pm0.6$ &\\
 \hline
\end{tabular}
\label{rot}
\end{table*}

\subsection{Residual velocities of young open clusters and model particles}

Residual velocities characterize  non-circular motions in the
Galactic disk.  We calculated them as   the differences between
the observed heliocentric velocities and the computed velocities
due to the circular rotation law and the adopted components of the
solar motion defined by the parameters listed in Table~\ref{rot}
(first line). For model particles the residual velocities are
determined with respect to the model rotation curve. We consider
the residual velocities in the radial $V_R$ and azimuthal $V_T$
directions. Positive radial residual velocities $V_R$ are directed
away from the Galactic center, while positive azimuthal residual
velocities $V_T$ are in the sense of Galactic rotation.

The velocity field of gas clouds moving along the outer elliptic
galactic rings is characterized by alternation of the negative and
positive radial residual velocities $V_R$ directed toward and away
from the galactic center, respectively
\citep{melnikrautiainen2009}.   In the ascending segments of the
rings (trailing spiral fragments) gas clouds have positive
velocities $V_R$ directed away from the galactic center, while in
the descending segments (leading spiral fragments), on the
contrary, $V_R$ is directed toward the center.   This becomes
clear when we remember that outer rings lie outside the CR of the
bar. Hence, in the reference frame co-rotating with the bar gas
clouds  rotate along the outer rings in the sense opposite that of
galactic rotation,  i.e. move in the sense of  decreasing
azimuthal angle $\theta$. In the descending segments, as defined,
galactocentric distance $R$ increases with increasing azimuthal
angle $\theta$ (leading spiral arm fragments), but  if objects
rotate in the sense of decreasing  angle $\theta$ the distance $R$
must decrease. So in the descending segments  of the outer rings
objects must approach the galactic center. In the 3 kpc solar
neighborhood there is only one descending segment of the outer
rings -- that of the ring $R_2$. Hence within 3 kpc of the Sun,
objects with negative velocities $V_R$  must outline the location
of the ring $R_2$.

\begin{figure*}
\resizebox{16 cm}{!}{\includegraphics{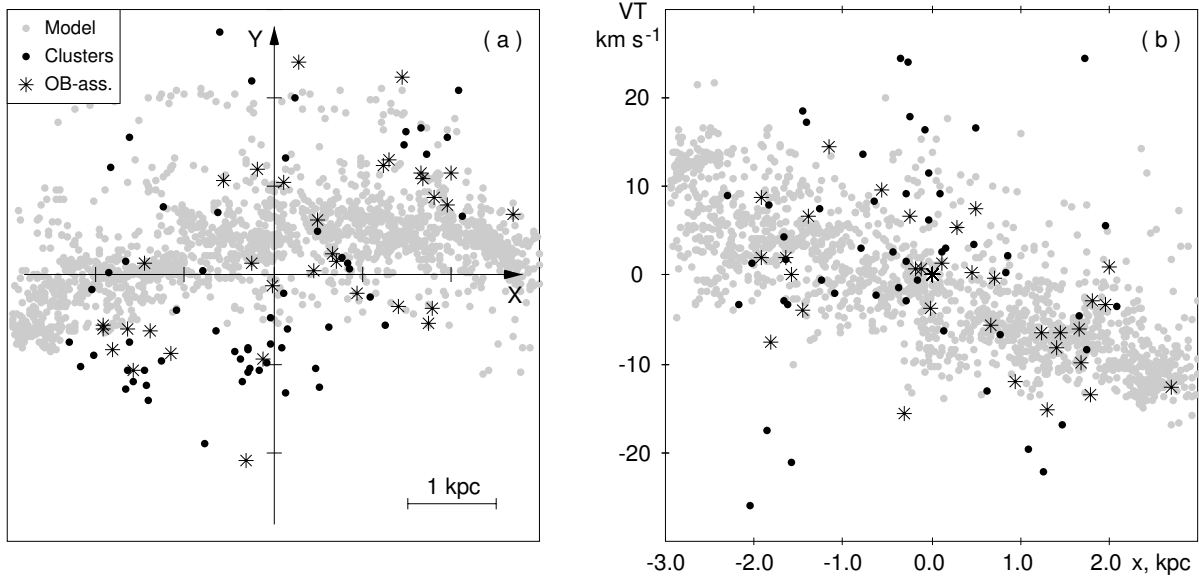}} \centering
\caption{Comparison of the distribution of  young open clusters
($\log \textrm{age} <8.00$) (the circles colored blue in
electronic edition) and OB-associations (the asterisks colored
green in electronic edition) with that of model particles (gray
circles). We selected only objects with radial residual velocities
directed toward the Galactic center ($V_R<0$). The positions of
model particles (gas and OB particles) are calculated for the
solar position angle $\theta_\textrm{b}=45^\circ$. (a)
Distribution of observed objects and model particles in the
Galactic plane. The selected clusters and OB-associations must
outline the descending segment of the outer ring $R_2$, i. e.
concentrate to a fragment of a leading spiral. The $X$-axis is
directed in the sense of the Galactic rotation and the $Y$-axis,
away from the Galactic center. The Sun is at the origin.  (b)
Dependence of azimuthal residual velocity $V_T$ on coordinate $x$
for objects with $V_R<0$. All objects demonstrate the decrease in
$V_T$ with increasing $x$.} \label{Vr_neg}
\end{figure*}

Figure~\ref{Vr_neg}a  shows the distribution of young open
clusters ($\log \textrm{age} <8.00$) and OB-associations with
negative radial residual velocities ($V_R<0$) in the Galactic
plane.  The sample includes 90 objects (57 clusters and 33
OB-associations) located within 3 kpc from the Sun. We consider
only OB-associations with the line-of-sight velocity and proper
motion based  on data  for  at least two cluster members
($n_{vr}\ge 2$, $n_{pm}\ge2$), and the same is true for open
clusters. Within $r<3.0$ kpc from the Sun, the elliptic ring $R_2$
can be represented as a fragment of  the leading spiral arm.  We
solve 90 equations by $\chi^2$-minimization \citep{press1987} to
find the parameters of the spiral law. The pitch angle of the
spiral-arm derived from clusters and OB-associations appears to be
$i=20\pm 5^\circ$. The positive value of the pitch angle $i$
indicates that Galactocentric distance $R$ increases with
increasing azimuthal angle $\theta$ what corresponds to the
leading spiral arm fragment and suggests the solar position near
the descending segment of the outer ring. Both types of objects,
young clusters and OB-associations, give the same result:
$i=19.3\pm 6.7^\circ$ and $i=20.3\pm 6.4^\circ$, respectively.
Obviously, these objects demonstrate similar behavior.

Note that the  pitch angle $i=20\pm 5^\circ$ obtained for clusters
and OB-associations differs strongly from the value derived for
model particles $i=6.6\pm 0.6^\circ$. The cause of this
discrepancy is that  objects of the Perseus and Carina complexes
deviate from the smooth contour of the outer ring $R_2$. If we
exclude the most  distant parts of these regions, for example, by
reducing the neighborhood considered from 3 to 2 kpc from the Sun,
the pitch angle estimate decreases from $i=20\pm 5^\circ$ to
$i=11.4\pm 5.3^\circ$. The  latter value was derived for 60
objects (39 clusters and 21 OB-associations) located within $r<2$
kpc of the Sun. The values of $i=11.4\pm 5.3^\circ$ and $i=6.6\pm
0.6^\circ$ obtained for observed objects and model particles are
consistent within the errors. The value of $i=11.4\pm 5.3^\circ$
is positive at the significance level $P>2\sigma$,  which means
that the spiral fragment to which young open clusters and
OB-associations concentrate is leading.

In this context the  question of the  ragged structure of the ring
$R_2$ comes up.  Though the deviations  from the smooth contour of
the ring $R_2$ can  result from errors in  the parameters of
distant objects, there is another aspect of the problem. A
two-component outer ring $R_1R_2$ usually  includes not just a
pure $R_2$ ring but a pseudoring $R_2'$ (broken ring)
\citep{buta1995}. Such deviations from  pure ring morphology are
also  observed in numerical simulations \citep[e.g.][Figure
3]{rautiainen2000} with the break  usually located on the
descending segment of the ring. Hence the fact that the Perseus
and Carina complexes  deviate from the smooth elliptic segment in
different directions -- away from the Galactic center in the
Perseus stellar-gas complex and toward the center in the Carina
complex -- may indicate the pseudoring morphology. This question
requires further study.

The residual azimuthal velocity $V_T$ also oscillates   along the
elliptic ring. These oscillations are shifted by $\pi/4$ in phase
with respect to those of  the radial velocity $V_R$ and are
associated with ordered epicyclic motion of gas clouds and young
stars near the OLR of the bar \citep{melnikrautiainen2009}.

Figure~\ref{Vr_neg}b shows the dependence of azimuthal residual
velocity $V_T$ on  coordinate $x$  for young open clusters,
OB-associations and model particles with negative radial residual
velocities ($V_R<0$). These objects are supposed to belong to the
descending segment of the outer ring $R_2$ and must show the
decrease of $V_T$ with increasing $x$. The regression coefficient
calculated for young clusters and OB-associations located within 3
kpc of the Sun is $k=-1.85\pm0.90$ and  differs significantly from
the value calculated for model particles $k=-3.45\pm0.07$.
However, if we leave only  OB-associations  we would get
$k=-3.03\pm0.85$, which is negative at significance level of
$P>3\sigma$ and agrees well with the model value.

\section{Discussion and conclusions}

We study the distribution and kinematics of young open clusters
from the catalogue by \citet{dias2002} in terms of the model of
the Galactic ring $R_1R_2'$ \citep{melnikrautiainen2009}. The best
agreement between the distribution of observed clusters and model
particles is achieved for the  solar position angle of
$\theta_\textrm{b}=35\pm10^\circ$ with respect to the bar major
axis.  It is  due to of "the tuning-fork-like" structure in the
distribution of clusters: at negative x-coordinates  most of the
clusters concentrate to the only one arm (the Carina arm), while
at positive x-coordinates  most of the clusters lie near the
Perseus or the Sagittarius complexes (Fig.~\ref{distrib_local}).

We studied the influence of objects located in  different regions
on the position  of the minimum of the $\chi^2$ curve. Excluding
the  clusters lying within the 0.5-kpc region ($r<0.5$ kpc) from
the observed sample decreases the value of $\theta_{\textrm{min}}$
from 35 to $30^\circ$, while the exclusion of distant clusters
located in the direction of the anticenter ($y>2.5$ kpc) increases
$\theta_{\textrm{min}}$ from 35 to $40^\circ$.

We performed mirror reflection of the distribution of observed
clusters with respect to the $Y$-axis, which is identical to the
($l \to 360^\circ -l$) transformation. The reflection obliterates
the minimum of the $\chi^2$ curve. After the reflection clusters
form a tuning-fork-like structure pointed in the opposite
direction (one segment at the positive $x$-coordinates and two
segments at the negative $x$-coordinates), which is inconsistent
with the position of the outer rings calculated for
$\theta_\textrm{b}=15\textrm{--}45^\circ$.

We simulated the influence of selection effects by assigning to
every model object the probability $P$ of its detection depending
on its apparent distance modulus $DM$.  The probability $P$ is
assumed to be equal to unity for $DM<DM_0$  and decrease
exponentially with $DM$ at $DM \ge DM_0$ (Eq.~\ref{p}). We
determined the parameters of the dependence $P(DM)$  by fitting
the distributions of distance moduli $DM$ of the observed and
model clusters (Figs.~\ref{his2},~\ref{min_par}). We computed the
extinction $A_V$ for model objects  in accordance with the
extinction of observed clusters located in the nearby region. We
scattered the model clusters in the vicinity of the model
positions of the outer rings and computed the  $\chi^2$ function
for 200 model samples. A comparison of the $\theta_\textrm{tru}$
and $\theta_\textrm{min}$ values reveals a small bias
$\Delta_s=-2.5^\circ$ which does not exceed the random errors of
$\pm5^ \circ$ (Table~\ref{chi2mod}).

We  consider  the value of $\theta_\textrm{b}=35^\circ$ as a good
compromise reflecting the combined  influence of different
effects. The upper limit of the combined error including random
and systematic errors is   $\pm 10^\circ$.

An analysis of the surface density $n$  of young clusters in
different sectors of the Galactic plane  suggests  a weak
dependence of  selection effects on the direction.  We  found that
extinction $A_V$ toward open clusters  grows most rapidly in the
direction of the Galactic center ($l=315\textrm{--}360^\circ$ and
0--45$^\circ$) and in the sector 45--90$^\circ$, while it is
minimal in the direction of the the Carina complex
270--315$^\circ$ and the anticenter 225--270$^\circ$. Note that we
found similar features in the variations of  extinction for
classical Cepheids \citep{melnik2015}. Possibly,  the sharp growth
of extinction  can be attributed to the presence of the Galactic
outer ring $R_1$ at the  heliocentric distance 1--2 kpc  in the
direction to the Galactic center.

The parameters of the rotation curve derived from the sample of
young open clusters  agree well with those obtained for
OB-associations \citep[][Table~\ref{rot}]{melnikdambis2009}. The
rotation curve is nearly flat in the 3 kpc solar neighborhood with
the large value of the Galactic angular velocity at the solar
radius $\Omega_0=30.3\pm1.2$ km s$^{-1}$ kpc$^{-1}$.

We study the distribution of  young open clusters and
OB-associations  with negative radial residual velocities $V_R$,
which within 3 kpc of the Sun must outline the descending segment
of the ring $R_2$. Clusters and OB-association demonstrate similar
distribution in the Galactic plane: objects of both types
concentrate to the fragment of the leading spiral arm \citep[see
also][]{melnik2005}. Within 2 kpc from the Sun, the pitch angles
of the spiral fragment derived for model particles
$i=6.0\pm0.5^\circ$ and observed objects $i=11.4\pm5.3^\circ$ are
consistent within the errors.

We  also found    the azimuthal velocity $V_T$  to decrease with
increasing  coordinate $x$ for objects with negative radial
residual velocities ($V_R<0$). The regression coefficient
calculated for 33 OB-associations located within 3 kpc of the Sun,
$k=-3.03\pm0.85$,  agrees  well with the value $k=-3.45\pm0.07$
calculated for model particles.

The morphological and kinematical features discussed  have also
been  found for the sample of classical Cepheids
\citep{melnik2015}. Thus, all types of objects -- young open
clusters, OB-associations and classical Cepheids --  suggest the
presence of the outer ring $R_1R_2'$ in the Galaxy.

\begin{acknowledgements}

We thank H. Salo for sharing his N-body code.  We are grateful to
O.~K.~Sil'chenko and A.~S.~Rastorguev for useful discussion. This
work was supported in part by the Russian Foundation for Basic
Research (project no 13\mbox{-}02\mbox{-}00203,
14\mbox{-}02\mbox{-}00472). Analysis of   open cluster data was
supported by Russian Scientific Foundation grant no. 14-22-00041.

\end{acknowledgements}

\end{document}